\documentclass[12pt]{article}
\newwrite\ffile\global\newcount\figno \global\figno=1

\def\writedef#1{}

\input epsf
\def\figin{\epsfcheck\figin}\def\figins{\epsfcheck\figins}
\def\epsfcheck{\ifx\epsfbox\UnDeFiNeD
\message{(NO epsf.tex, FIGURES WILL BE IGNORED)}
\gdef\figin##1{\vskip2in}\gdef\figins##1{\hskip.5in}% blank space instead
\else\message{(FIGURES WILL BE INCLUDED)}%
\gdef\figin##1{##1}\gdef\figins##1{##1}\fi}

\def\figinsert{}
\def\ifig#1#2#3{\xdef#1{fig.~\the\figno}
\writedef{#1\leftbracket fig.\noexpand~\the\figno}%
\figinsert\figin{\centerline{#3}}\medskip\centerline{\vbox{\baselineskip12pt
\advance\hsize by -1truein\center\footnotesize{  Fig.~\the\figno.} #2}}
\bigskip\endinsert\global\advance\figno by1}
\def\endinsert{}

\begin{document}
\baselineskip 18pt
\newcommand{\Tr}{\mbox{Tr\,}}
\newcommand{\beq}{\begin{equation}}
\newcommand{\eeq}[1]{\label{#1}\end{equation}}
\newcommand{\bea}{\begin{eqnarray}}
\newcommand{\eea}[1]{\label{#1}\end{eqnarray}}
\renewcommand{\Re}{\mbox{Re}\,}
\renewcommand{\Im}{\mbox{Im}\,}
\begin{titlepage}

\begin{picture}(0,0)(0,0)
\put(350,0){SHEP-00-05}
\put(350,-15){Imperial/TP/99-0/28}
\end{picture}
 
\begin{center}
\hfill
\vskip .4in
{\large\bf AdS RG-Flow and the Super-Yang-Mills Cascade}
\end{center}
\vskip .4in
\begin{center}
{\large Nick Evans$^{a}$ and  Michela Petrini$^b$}
\footnotetext{e-mail: n.evans@hep.phys.soton.ac.uk, m.petrini@ic.ac.uk }
\vskip .1in
(a){\em Department of Physics, Southampton University, Southampton,
S017 1BJ, UK}
\vskip .1in
(b){\em Theoretical Physics Group, Blackett Laboratory,
Imperial College, London SW7 2BZ, U.K.}
\end{center}
\vskip .4in
\begin{center} {\bf ABSTRACT} \end{center}
\begin{quotation}
\noindent We study the 5 dimensional SUGRA AdS duals of N=4, N=2 and N=1
Super-Yang-Mills theories. To sequentially break the N=4 theory mass terms
are introduced that correspond, via the duality, to scalar VEVs in the SUGRA.
We determine the appropriate scalar potential and study solutions
of the equations of motion that correspond to RG flows in the field theories.
Analysis of the potential at the end of the RG flows distinguishes the flows
appropriate to the field theory expectations. As already identified in the 
literature, the dual to the 
N=2 theory has flows corresponding to the moduli space of the field theory. 
When the N=2 theory is broken to N=1 the single flow corresponding to the 
singular point on the N=2 moduli space is picked out as the vacuum. As the 
N=2 breaking mass scale is increased the vacuum deforms smoothly to previously
analysed N=1 flows.

\end{quotation}
\vfill
\end{titlepage}
\eject
\noindent
\section{Introduction}

The duality between weakly coupled
string theory on $AdS_5 \times S_5$ and large $N$,
N=4 super-Yang-Mills (SYM) theory \cite{malda,gkp,w1} is by 
now well established. The correspondence,
that excitations in the string/SUGRA theory act as sources for operators
in the field theory at the boundary,  suggests that mass terms
in the field theory will correspond to VEVs of scalar fields in the $AdS$
space.
Several authors have explored SUGRA duals to theories which in the IR
correspond to softly broken N=4 in this fashion \cite{gppz1}-\cite{bs2}.
Many of those results have been obtained by reduction to five dimensions
where the problem can be studied in terms of a theory of scalars coupled
to gravity, and we will pursue this approach further in this paper. 

The first field theories studied in this 
way had both IR and UV fixed points which in SUGRA corresponded to
solutions of the relevant scalar equations of motion that flowed between
two fixed points of the scalar potential in the radial direction in AdS
($y$) \cite{gppz1} - \cite{lust},\cite{freed1}. 
Following the interpretations in \cite{malda,PP} 
the functional dependence on $y$ corresponds to RG 
flow in the field theory. Flows that at large $y$ approach
the N=4 fixed point at the origin of the scalar potential but at small
$y$ flow down the potential to a singularity have also been studied.
They are interpreted as RG flows in a theory with no IR fixed point (at low
scales the mass terms grow without bound). In this fashion SUGRA duals 
of strongly coupled relatives of 
N=2 and N=1 theories have been studied \cite{freed1},\cite{gppz3}-\cite{bs2}. 
It must always be remembered that 
introducing mass term perturbations to the N=4 field theory does not decouple
the massive fields from the strong dynamics because the N=4 theory is conformal
and strongly coupled above the breaking scale. Supersymmetry hopefully
ensures that the resulting theories do lie in the same universality class as their
cousins with weak coupling at the breaking scale though. 

The singularities in the 5$d$ SUGRA                             
approach to non-conformal field theories  show that in the deep IR a fuller stringy picture of the dynamics
is required and a number of papers have begun to study these descriptions
(the singularities hide such objects as the enhancon singularity in N=2
\cite{pjp} and the Myers' D-brane polarisation effect in N=1
\cite{my,ps}). Nevertheless as we will see the SUGRA 
description still contains much of the physics of the field theories. 

A set of flows have been identified
in the SUGRA dual to N=2 SYM \cite{gub2,pw,bs2} and some criteria must be used to distinguish
between different flows to identify the physical ones. In \cite{gub2} it
has been proposed that the potential evaluated along physical flows should
remain bounded by the asymptotic value of the potential at the origin. 
Imposing this condition, as we discuss further below, picks out a set of
flows that reasonably correspond to the RG flows in the field theory 
associated with different choices of position on the scalar moduli space.
The extremum flows are the natural candidates to be identified with 
the singular points on the N=2 theory's moduli space. In \cite{gppz3}
flows corresponding
to N=1 SYM were investigated. With the introduction of an appropriate
mass term (scalar) a single flow was found (up to the ability to rescale
the RG parameter $y$ and neglecting the gaugino condensate). 
In this paper we want to explore the set of models
that lie between these two extremes. The field theory of the N=2 model 
is known to have a quantum mechanical moduli space which at a general point has
a $U(1)^{N-1}$ gauge symmetry in the IR \cite{sw1}. 
The couplings of the U(1) gauge fields are
determined by the periods of the Seiberg-Witten curve \cite{sw1}
\beq
y^2 = \prod_{i=1}^N (x - \phi_i)^2 - \Lambda^{2N},
\eeq{N=2curve}

The singular points at $tr \phi_i^2 = \Lambda^2$ correspond to places
where the U(1) couplings diverge and there are massless dyons charged
under the U(1)s. When the theory is perturbed by the addition of a mass
term for the scalar multiplet the resulting potential pins the theory
at the singular points. Holomorphy determines that as the mass terms are 
increased the vacuum must smoothly deform to the vacuum of N=1 SYM 
with the scalar VEV approaching zero. On the field theory side, 
the breaking N=4 $\rightarrow$ N=2 $\rightarrow$ $N=1$ has been studied
in \cite{cas}. In the analysis below we will 
present the SUGRA dual of this pinning on the moduli space and evolution
to the N=1 SUGRA flow with no field theory scalar VEV.

\section{Mass Terms and SUGRA Scalar Potentials} 

Our starting point is the N=4 SYM theory which, in N=1 language, has
three, adjoint  chiral multiplets ($\Phi_1, \Phi_2, \Phi_3$) in addition to the gauge multiplet.
Previously deformations of the N=4 theory with an equal mass term for two
of the gauge chiral multiplets \cite{gub2,pw,bs2} 
and equal mass terms for all the three N=1 chiral multiplets 
\cite{gppz3} have been considered.
Here we will consider a generalisation with different 
masses for the chiral superfields. More precisely, we give equal mass, $m$ to two 
of the three superfields, and mass $M$ to the other one.  
In N=1 notations, this corresponds to
\beq
W= m \sum_{i=1,2}\Phi_i^2 + M \Phi_3^2
\eeq{m21}
where $m$ and $M$ are complex.

For $m=M$, this deformation corresponds to N=4 SYM softly broken to N=1 
\cite{gppz3}, while for $M=0$ one recovers N=4 SYM softly broken to N=2 
\cite{gub2,pw,bs2}.
For $M<<m$, our N=1 solution should correspond to the soft breaking of 
N=2 to N=1 described above. As we increase $M/m$ we should smoothly return to
the N=1 solutions.

In the SUGRA description of the N=4 theory the mass terms (sources)
in the field theory correspond to VEVs of the SUGRA fields. The
precise fields have been identified from their symmetry properties
under the conformal group (which indicates they are scalars) 
and the $SU(4)_R$ global symmetry of the N=4
theory. 
We will now identify the appropriate scalars and construct the five-dimensional
supergravity 
solution corresponding to the deformation (\ref{m21}).
The scalars of N=8 gauged supergravity  are in the coset 
$E_6/USp(8)$ \cite{warn}. 
The elements of the coset are $27\times 27$ matrices, $U$, transforming in the 
fundamental representation of $E_6$. In a unitary gauge, $U$
can be written as $U=e^X, X=\sum_a \lambda_a T_a$,
where $T_a$ are the
generators of $E_6$ that do not belong to $USp(8)$. 
This matrix is parametrised by 42 real scalars, which are the physical scalars 
of the theory. They transform as the 
${\bf 10}+ \bar{\bf 10}$, ${\bf 20'}$, and ${\bf 1_c}$ of the gauge 
group $SO(6)$.
The singlet is associated with the marginal deformation
corresponding to a shift in the coupling constant of the N=4 theory.
The mode in the ${\bf 20'}$ is
associated with a symmetric traceless mass term for the scalars
\beq
 \Tr(X_IX_J)-\frac{1}{6} \delta_{IJ} Tr(X_LX_L), \quad I,J=1,\ldots,6,
\eeq{m22}
while the ${\bf 10} + \bar{\bf 10}$ correspond to 
the fermion mass term 
\beq
\Tr\lambda_A\lambda_B + h.c., \quad A,B=1,\ldots4.
\eeq{m23}

A generic supersymmetric mass term for the three chiral multiplets
corresponds to turning on scalars both in the ${\bf 10} + \bar{\bf 10}$ and 
${\bf 20'}$.
Let us consider first the fermion mass term:
$m_{ij} \Tr(\lambda_i \lambda_j)$ with $i=1,2,3$.
$m_{ij}$ is a complex, symmetric matrix that
transforms as the ${\bf 6}$ of $SU(3)\in SU(4)$ ($SO(6)$ $\sim
SU(4)$). The corresponding
supergravity mode appears in the
decomposition of the ${\bf 10}\rightarrow
{\bf 1}+ {\bf 6} + {\bf 3}$ of $SU(4)$ under
$SU(3)\times U(1)$. 
The singlet  in this  
decomposition corresponds to the scalar $\sigma$ dual to the gaugino
condensate of the  N=1 SYM. In the analysis to follow we will set 
this field to zero in order to simplify the  computation of the potential
and reduce the dimension of the parameter space of flows. In principle
it should be present and is presumably non-zero though, as we will see, 
neglecting it  does not appear to disrupt the physical interpretation
of flows in the remaining fields. 

In principle, a non-zero VEV for $m_{ij}$ will
induce non-zero VEVs for other scalars as well, due to the existence
of linear couplings of $m$ to other fields in the potential.
A simple group theory analysis shows that the only couplings 
of the ${\bf 6}$ that give rise to a singlet of the symmetry 
group are of the form ${\bf 8}\times 
(\bar{\bf 6} \times {\bf 6})^k$ where $k$ is an integer number 
and the ${\bf 8}$ appears in the decomposition of the  
 ${\bf 20}'\rightarrow
{\bf 8}+ {\bf 6} + \bar{\bf 6}$ of $SU(4)$ under
$SU(3)\times U(1)$. 
The ${\bf 8}\subset {\bf 20}'$ is then the only other mode that has to be 
considered, and all the remaining fields
can be consistently set to zero. 
%This is true also if we consider a two-parameter Lagrangian depending on both
%$m$ and $\sigma$.

Notice that the ${\bf 8}$ corresponds exactly to the scalar mass term
one would  expect on the field theory side. 
Indeed by supersymmetry the mass term for the scalars is the square of the fermionic one, $\Lambda_{ij}=m_{ij}m^{*}_{jl}$ $\leftrightarrow$ 
${\bf 1}
+{\bf 8} \subset 
{\bf 6} \times \bar{\bf 6}$. The singlet, which amounts to the trace of
the scalar mass terms, is associated with a massive stringy state and has no dual 
SUGRA description (the SUGRA only contains the massless string sector).
An added complication is that a scalar VEV in the field theory has the 
same symmetry properties as the scalar mass term and is therefore
also described by the ${\bf 20'}$.

The  deformation of eq.(\ref{m21}) corresponds to taking the matrices 
$m_{ij}$ and $\Lambda_{ij}$ diagonal. In the complex basis of
\cite{gppz3}, they read\footnote{The factors of $\sqrt{2}$ and
$\sqrt{6}$  are required in supergravity
in order for the fields to have canonical kinetic term.}:  
\beq
(m_{ij})=\mbox{diag}\,(\frac{m}{\sqrt{2}},\frac{m}{\sqrt{2}},M) \qquad \mbox{and} \qquad 
(\Lambda_{ij})=\frac{\rho}{\sqrt{6}} \, \mbox{diag}\,(-1,-1,2).
\eeq{m24}
If we fix $m_{ij}$ then for $\Lambda_{ij}$ to correspond to the 
supersymmetric mass terms we must assume that the massive stringy mode
diag(1,1,1) has also developed a VEV though this is not explicit
in the SUGRA (the ability to describe these solutions must be implicitly
present in the SUGRA). For a given choice of $\rho$ we assume the
the stringy mode VEV brings the first two elements in line with 
supersymmetric requirements. We interpret the discrepancy in the third
element from 
$M^2$ as a VEV for $tr \phi_3^2$. Thus changing $\rho$,
with fixed $m_{ij}$, allows us to explore the N=2 theory's moduli space.
Note that $tr \phi_3^2$
is complex whilst $\rho$ is real, so we will only be able to explore the
moduli space along a single radial direction. In the large $N$ limit though 
the $Z_N$ symmetry is restored to a U(1) so the radial direction is
sufficient.

The lengthy computation of the potential
and kinetic terms is performed along the lines of \cite{gppz3}.
We refer 
to previous papers \cite{dz,pilch,grw2} for an extensive description of
these kinds of calculation. 
The 5-dimensional action \cite{grw2}
for the scalars $m, M$ and $\rho$ is\footnote{In the
following we will always set the coupling constant $g$ equal to $2$, so
that the scalar potential in the $N=8$ vacuum, where all scalars have
zero VEV, is normalised as $V(N=8)=-3$.}
\bea
L = \sqrt{-g}\{- {R\over 4}
+ {1\over 2}(\partial m)^2+{1\over 2}(\partial M)^2 
+ {1\over 2}(\partial \rho)^2 + V\},
\eea{m25}      
where the potential $V$ is given by
\bea
V &=& \frac{1}{8}e^{\frac{-4\rho}{\sqrt{6}}}\left[-5+\cosh(4M)-4\cosh(2M)\right]
- e^{\frac{2\rho}{\sqrt{6}}}\cosh(\sqrt{2}m)\left[\cosh(2M)+1\right]+ 
\nonumber\\
& & + \frac{1}{16} e^{\frac{8\rho}{\sqrt{6}}}
\left[-3+2\cosh(2\sqrt{2}m)+\cosh(4M)\right].
\eea{m26}
The potential above contains as special case the examples previously
studied in the literature: for $M=0$  and $M=\sqrt{2}m$ it gives the N=2
\cite{pw,bs2} and 
N=1 \cite{gppz3,pz} potentials, respectively, while for $m=0$ it reduces
to the potential for the flow to the N=1 supersymmetric fixed point
described in \cite{freed1}.

In ref.~\cite{freed1,townsk} 
the conditions for a supersymmetric flow were found. 
For a supersymmetric solution, the potential $V$
can be written in terms of a superpotential $W$ as
\beq
V = \frac{1}{8} \sum_{a=1}^{n} \left|
\frac{\partial W}{\partial \lambda_a} \right|^2
- \frac{1}{3} \left|W \right|^2.
\eeq{27}
In terms of the computation of the potential as described in \cite{grw2} 
$W$ is one of the eigenvalues of the tensor $W_{ab}$.
Moreover, for such a supersymmetric solution, since the fermionic shifts
vanish,  the second order equations reduce to first order ones
\bea
\dot\lambda_a&=&\frac{1}{2} \frac{\partial W}{\partial \lambda_a},\\
\dot{\phi}&=& - \frac{1}{3} W.
\eea{28}

Here $\phi$ is the scale factor in the 5-dimensional metric
\beq
ds^2= dy^2 + e^{2\phi(y)}dx^\mu d x_\mu , \;\;\;\mu =0,1,2,3,
\eeq{29}
where $y$ is the fifth coordinate of $AdS_5$, which we interpret  as  
an energy scale \cite{malda,PP}:  $y \rightarrow \infty$ 
corresponds to the UV regime while $y\rightarrow - \infty$ to the IR.
The dot in eq.(\ref{28}) indicates derivative with respect to $y$.

The $W_{ab}$ tensor of \cite{grw2}, has two different eigenvalues, 
both with degeneracy 2, that satisfy the condition in eq.(\ref{27}),
\bea
W_1&=& -\frac{1}{2} e^{\frac{4\rho}{\sqrt{6}}}\left(2\cosh(\sqrt{2}m)
+\cosh(2M)- 1\right)
- e^{\frac{-2\rho}{\sqrt{6}}}\left(\cosh(2M)+1\right)\\
W_2&=& -\frac{1}{2} e^{\frac{4\rho}{\sqrt{6}}}\left(2\cosh(\sqrt{2}m)
-\cosh(2M)+ 1\right)
- e^{\frac{-2\rho}{\sqrt{6}}}\left(\cosh(2M)+1\right),
\eea{210}
indicating that there is more than one superpotential
that generates the potential.
We then expect to have two N=1 supersymmetric flows, with different field 
theoretic interpretations depending on the asymptotic behaviour of the
fields for $y\rightarrow \infty$ \cite{bala,kw2}. These are obtained by
substituting into the equations of motion the  
linearised expressions for the eigenvalues (\ref{28}) in the neighbour of the
N=4 fixed point ($m=M=\rho=0$): $W_1\sim
-3-m^2-3M^2-2\rho^2-\frac{4}{\sqrt{6}}\rho \,m^2$ and $W_2 \sim -3-m^2-M^2-2\rho^2-\frac{4}{\sqrt{6}}\rho \,
\left(m^2-2M^2\right)$.
Given those expressions, it is easy to check that 
while $m$ always scales as a mass deformation ($m \sim e^{-y}$)
\cite{bala,kw2}, the behaviour of $M$ depends on the choice of $W_i$: 
a scalar VEV $M \sim
e^{-3y}$ for $W_1$ or a
mass deformation $M \sim e^{-y}$ for $W_2$. 
Thus flows of softly broken N=2 SYM should correspond to solutions of
the equation of motion with $W=W_2$.
Notice also that the same eigenvalue $W_2$ gives the equations of
motions for the N=1 and N=2 cases mentioned above.

\section{Properties of the solutions}

As discussed in the previous section, the SUGRA duals of RG flows of
softly broken N=2 super Yang-Mills should be given by solutions of the
equations  
\bea
\dot{\phi}&=& \frac{1}{6} e^{\frac{4\rho}{\sqrt{6}}}\left(2\cosh(\sqrt{2}m)
-\cosh(2M)+ 1\right)
+\frac{1}{3} e^{\frac{-2\rho}{\sqrt{6}}}\left(\cosh(2M)+1\right),\\
\dot{\rho}&=& -\frac{1}{\sqrt{6}}\left[e^{\frac{4\rho}{\sqrt{6}}}\left(2\cosh(\sqrt{2}m)
-\cosh(2M)+ 1\right)
- e^{\frac{-2\rho}{\sqrt{6}}}\left(\cosh(2M)+1\right)\right],\\
\dot{m}&=& - \frac{1}{\sqrt{2}} e^{\frac{4\rho}{\sqrt{6}}}\sinh(\sqrt{2}m),\\
\dot{M}&=& \frac{1}{2}\left(e^{\frac{4\rho}{\sqrt{6}}}
-2 e^{\frac{-2\rho}{\sqrt{6}}}\right) \sinh(2M),
\eea{31}
with the boundary conditions that the scalars $m,M$ and $\rho$ vanish
and $\phi \sim y$ for $y\rightarrow \infty$ .

Contrary to the N=1 and N=2 cases, these equations do not seem to be
analytically solvable, and we must rely upon numerical results.
The solutions will depend on a certain number of parameters 
which correspond to the value of the fields at a given UV scale. 
The asymptotic behaviours of the fields for $y\rightarrow
\infty$ are
\bea
m &\sim&  m_{0} e^{-y},\\
M &\sim&  M_{0} e^{-y},\\
\rho &\sim & -\sqrt{2/3}\left(m_{0}^2-M_{0}^2 \right) e^{-2y} y +
\rho_{0}e^{-2y},\\ 
\phi &\sim& \phi_{0}+ y.  
\eea{32}
The coefficients $m_0$ and $M_0$ corresponds to the UV field theory masses
for the fermions. As expected, we can distinguish two contributions in 
the asymptotics of $\rho$: the first one corresponds to a mass term for
the scalar components of $\Phi_{1,2}$ and indeed it  has the coefficient fixed by supersymmetry in
terms of the fermion masses, the second term is associated to a VEV for the
scalar of $\Phi_3$ and the arbitrary coefficient represents the freedom in moving
along the moduli space of the theory. 
In our numerical analysis we will not distinguish between these two
different contributions, and we will just indicate the UV values of the
fields as $M(y_{UV})$, etc...

We first consider the N=2 ($M=0$) case for which an analytical solution
has been given \cite{pw}. This provides us with an interesting check of
the numerical analysis, that we can then extend to the N=1 solution. 

\subsection{N=2 Flows}
Setting $M=0$ we find the N=2 flows described by the equations
\bea
\dot{\rho} &=& - {2 \over \sqrt{6}}  \left
(e^{4 \rho /\sqrt{6}} \cosh \sqrt{2} m
- e^{-2 \rho /\sqrt{6}} \right), \\
\dot{m} &=& - {1 \over \sqrt{2}} e^{4 \rho / \sqrt{6}} \sinh \sqrt{2} m,
\eea{meep2}
from which we find
\beq
{\partial \rho \over \partial m} = {2 \over \sqrt{3} \sinh \sqrt{2}m} 
\left( \cosh{\sqrt{2}m} - e^{-6 \rho /\sqrt{6}} \right).
\eeq{meep3}
The solution to this equation was given in \cite{pw}
\beq 
e^{\sqrt{6} \rho} = \cosh \sqrt{2} m  + \sinh^2 \sqrt{2} m
 \left[ k + \log \left(
{\sinh m/ \sqrt{2} \over \cosh m/ \sqrt{2} }  \right) \right].
\eeq{meep4}
The constant $k$ that parametrises the solutions is related to the
UV values of the fields by 
$ \left(2k +1
-\log{2}\right)\sqrt{6}= \rho_0/m_0^2 -\sqrt{2/3} \log{m_0} $.  

\begin{figure}
\epsfxsize 10cm \centerline
{\epsffile{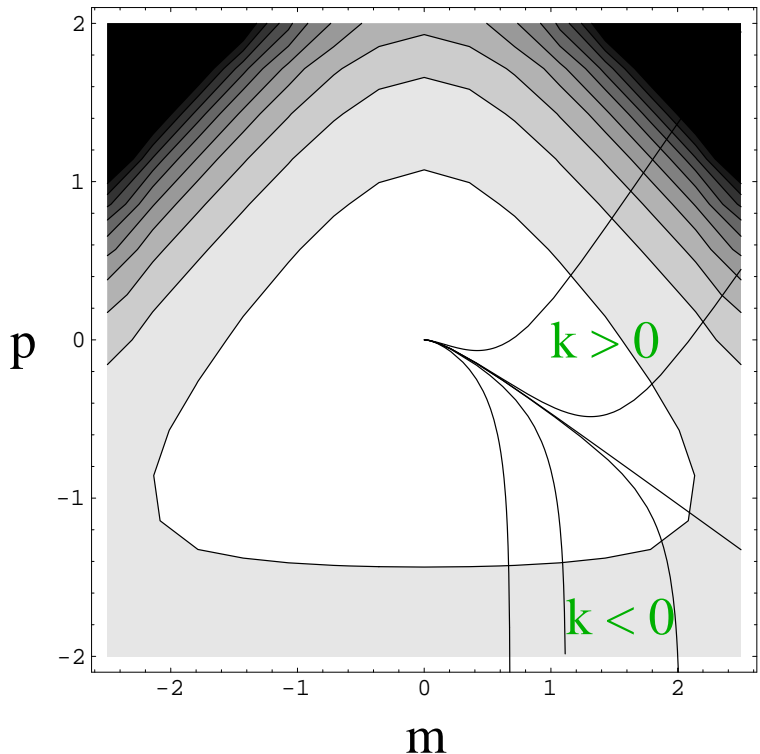}}
\caption{Solutions of the equation of motion for the N=2 SUGRA dual
superimposed on the value of the superpotential (lighter regions
correspond to larger values) in the $\rho-m$ plane.
  }
\label{Fig1}
\end{figure}

We plot these flows over the superpotential in Fig. \ref{Fig1}.
In order to interpret these flows in terms of the N=2 gauge theory 
one requires some criteria for distinguishing between acceptable
and unacceptable flows. In \cite{gub2} it has been proposed that the
flows may be 
distinguished based on the behaviour of the potential and superpotential
along the flows. In particular, extrapolating from finite temperature
situations where there are requirements for consistent black hole
solutions, the author has proposed that only flows where the scalar
potential is bounded above all along the flow are physical. 
This seems intuitive since these are
flows that begin at the origin of the $\rho -m$ plane at large $y$ 
(that is they look like the N=4 theory in the UV) and then flow away
from the origin, in directions where the potential falls,
to large VEVs at small $y$ (the field theory IR). 

\begin{figure}[ht]
\epsfxsize 17cm \centerline
{\epsffile{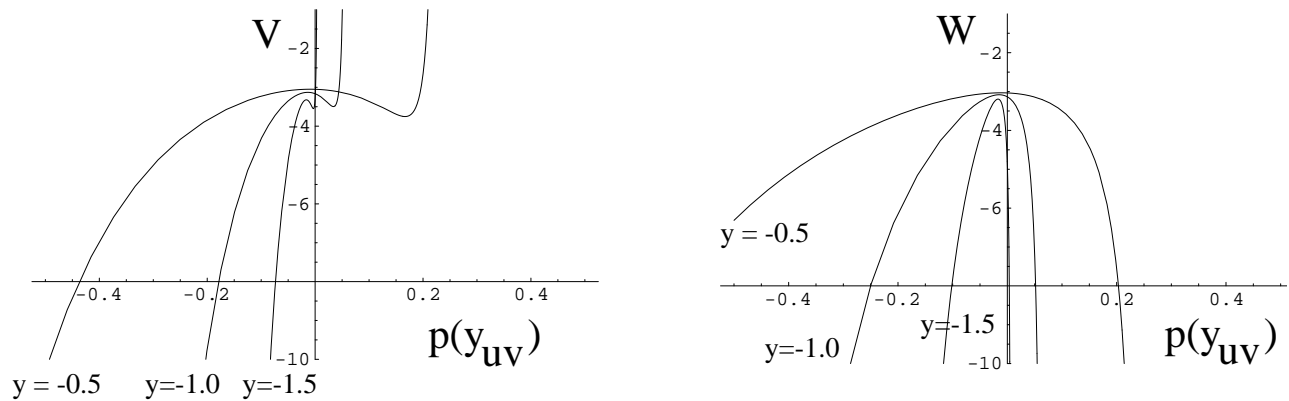}}
\caption{The potential and superpotential evaluated at varying $y$ along
a set of flows in the N=2 theory starting at $y_{UV}=0.1$ with $m=0.1$ and 
varying $\rho(y_{UV})$.}
\label{Fig2}
\end{figure}

To discuss this criteria for the flows of Fig.\ref{Fig1} it is sensible
to formulate boundary conditions that are easily interpretable in the
field theory. In the field theory it is natural to start at some
UV scale with a fixed mass and look for RG flows that correspond to
different positions on the moduli space of the theory. Thus in the SUGRA
we should fix $m$ at some $y = y_{UV}$ and
look at flows with varying $\rho$. We are therefore taking initial conditions
on a vertical slice through Fig 1. We are then interested in the behaviour 
of the potential along the flow to lower $y$. We plot the evolution
of the potential (and superpotential) for several values of $y$ 
along such flows using numerical solutions of the field
equations in Fig.\ref{Fig2}. 

The behaviour is that all flows which start with $\rho(y_{UV})$ greater than 
the value on the $k=0$ ``ridge flow'' eventually meet a positive potential
wall, and hence by the  conditions in \cite{gub2}
any flows with $k > 0$ are to be considered unphysical. That
the $k=0$ curve is the critical curve where this behaviour ends is
clear from Fig 1 since only for the $k >0$ curves do
the flows reach the singular behaviour of the superpotential which
corresponds to that of the potential. 

It seems likely therefore that the flows with $k \leq 0$ correspond
to the RG flows of different points of the field theory moduli space.
There is some evidence to suggest this is true. For the curves $k < 0$, 
as can be seen in Fig 1, the flows asymptotically approach $\rho \rightarrow 
- \infty$ and $m \rightarrow$ constant. Evaluating the potential eq.(7) 
in this 
limit shows that the first term in the potential dominates and the 
potential becomes independent of $m$. Thus these flows asymptotically
see the same potential suggesting a moduli space. That the different flows
approach the asymptotic form of the potential
at different rates with respect to $y$ is perhaps an indication that in
the field theory different points on the moduli space have a different
scale $tr \phi_3^2$ at which the gauge symmetry is broken, $SU(N) \rightarrow U(1)^{N-1}$. 
The singular point on the N=2 moduli space, where the U(1) couplings diverge,
should 
naturally be equated with a special or extremal flow. 
The obvious such flow is the case
$k=0$ which follows the crest of the ridge on the superpotential
\cite{gub2,pw}. 
These identifications can only be tentative based on the analysis so far. 
In \cite{pw} further evidence was provided by lifting the 5$d$ SUGRA 
solution to a 10$d$ SUGRA solution. This allows the authors  to 
evaluate the gauge coupling (corresponding to the VEVs of the 
singlet scalars amongst the 42 scalars in the 5$d$ SUGRA theory - 
in the 5$d$ theory
they do not enter the potential and so their VEVs can not be determined).
They find that the coupling diverges on the $k=0$ flow but runs to
a constant elsewhere which seems in accord with  the field theory
although the functional dependence of the coupling on the moduli space
has not been reproduced.

We next move on to consider breaking the N=2 theory to the N=1 theory by the 
inclusion of the mass term $M$. In the field theory we expect to be pinned at 
the singular point and indeed we will see that the SUGRA flows pick
out the $k=0$ flow providing
further evidence for the above identification of the flows with the
N=2 moduli space.

\subsection{N=1 Flows}

\begin{figure}[ht]
\epsfxsize 17cm \centerline
{\epsffile{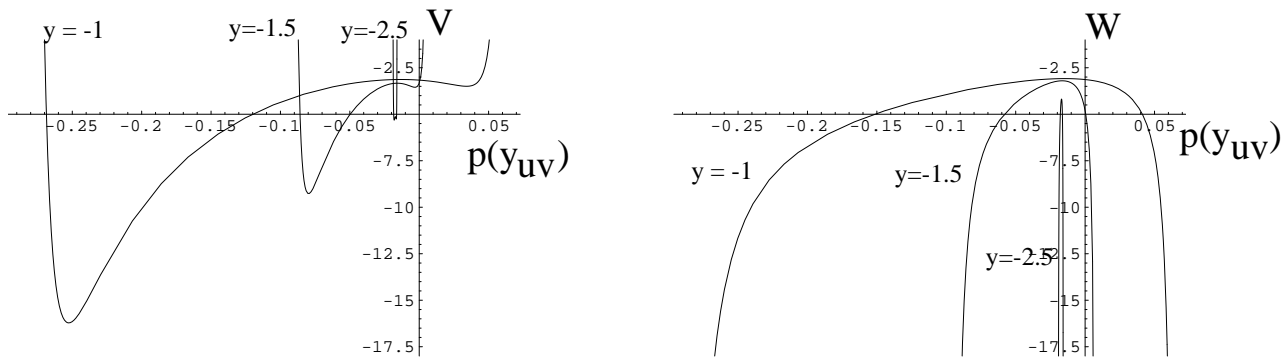}}
\caption{Plots of $V$ and $W$ at varying $y$ in the N=1 model with 
flows starting at $y_{UV}=0.1$ and $M(y_{UV})=0.1m(y_{UV})$ with
$m(y_{UV})=0.1$ and varying $\rho( y_{UV})$. The pinning of the potential 
on the N=2 critical flow is apparent as one approaches the IR.
  }
\label{Fig3}
\end{figure}

Flows with non-zero $M$ are expected to correspond to N=1 super-Yang-Mills
theories. We begin by looking at theories which are N=2 SYM plus a small 
mass $M$ in the UV that breaks the supersymmetry to N=1.  
We again, in the SUGRA, fix $m$ and $M$ at some $y = y_{UV}$ and
look at flows with varying $ \rho(y_{UV})$. We are then interested in
the behaviour  
of the potential along the flow to lower $y$. In Fig.\ref{Fig3} we plot
the evolution 
of the potential (and superpotential) for several values of $y$ 
along such flows using numerical solutions of the field
equations with $M(y_{UV}) = 0.1 m(y_{UV})$. 
From Fig.\ref{Fig3} it is apparent that as we proceed towards the IR
all the flows except the flow along the ridge of the superpotential
eventually meet a positive ``barrier'' in the potential. Thus
only the single ridge flow is allowed by the criteria for distinguishing 
flows discussed above. This matches our expectations from field theory
where we expected the theory to become pinned at the singular point of the N=2 
theory previously identified with that ridge flow.

\begin{figure}[ht]
\epsfxsize 15cm \centerline
{\epsffile{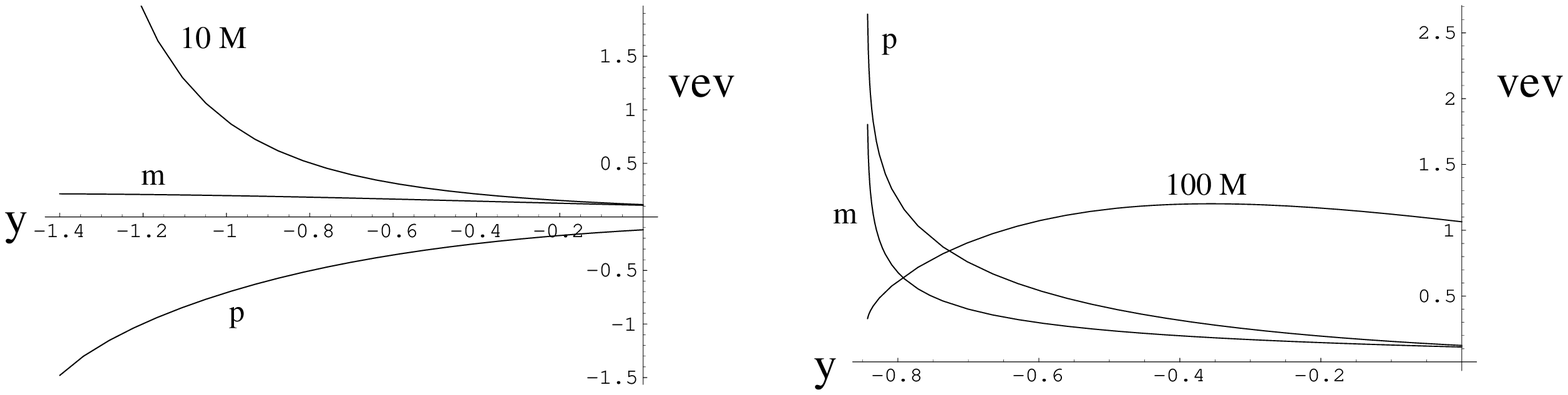}}
\caption{The evolution of the individual fields $\rho, m, M$ with $y$.
 The left hand plot corresponds to a $k <0$ flow, the right hand plot to $k>0$. Initial conditions are as in Fig 3 above.
  }
\label{Fig4}
\end{figure}

The existence of a ridge flow seems to be confirmed by the analysis of
the numerical solutions of the equations of motion.
One can indeed distinguish two sets of solutions with radically
different behaviours. 
The change of behaviour seems to take place for a negative value
of the $\rho(y_{UV})$, which should correspond to the position of the
ridge flow. This is consistent with what shown in Fig \ref{Fig3}.
In Fig. \ref{Fig4} we plot $M,m$ and $\rho$ as functions of $y$ for two
values of $\rho(y_{UV})$ on either sides of the ridge flow.
In analogy with the N=2 theory case, we will call the
ridge flow $k=0$, while
$k<0$ and $k>0$ will indicate the flows on the left and on the right of it,
respectively. Notice however that here the parameter $k$ is not related with
any precise form of the solutions.

Unfortunately it is not very easy to reproduce the  behaviour of the
ridge flow.  Numerically, it is very difficult to pick out the precise
value of $\rho(y_{UV})$ corresponding to the ridge flow,
and analytically, it is not easy to solve the equations of motion
even in the IR limit. It is even hard to guess the correct asymptotic
behaviour of the fields. For example, one would have hoped to recover a
logarithmic 
behaviour like in the N=1 or N=2 case, but this possibility seems to be
ruled out. This is confirmed by the plots in Fig. \ref{Fig5} below, where the
behaviour of the ratios of the various fields as a function of $y$ are
shown. The existence of, and distinction between, the ridge flow 
and the two classes
of flows to either side are clear, but note that close to
the ridge flow there is no linear relation between the
fields.
\begin{figure}[ht]
\epsfxsize 15cm \centerline
{\epsffile{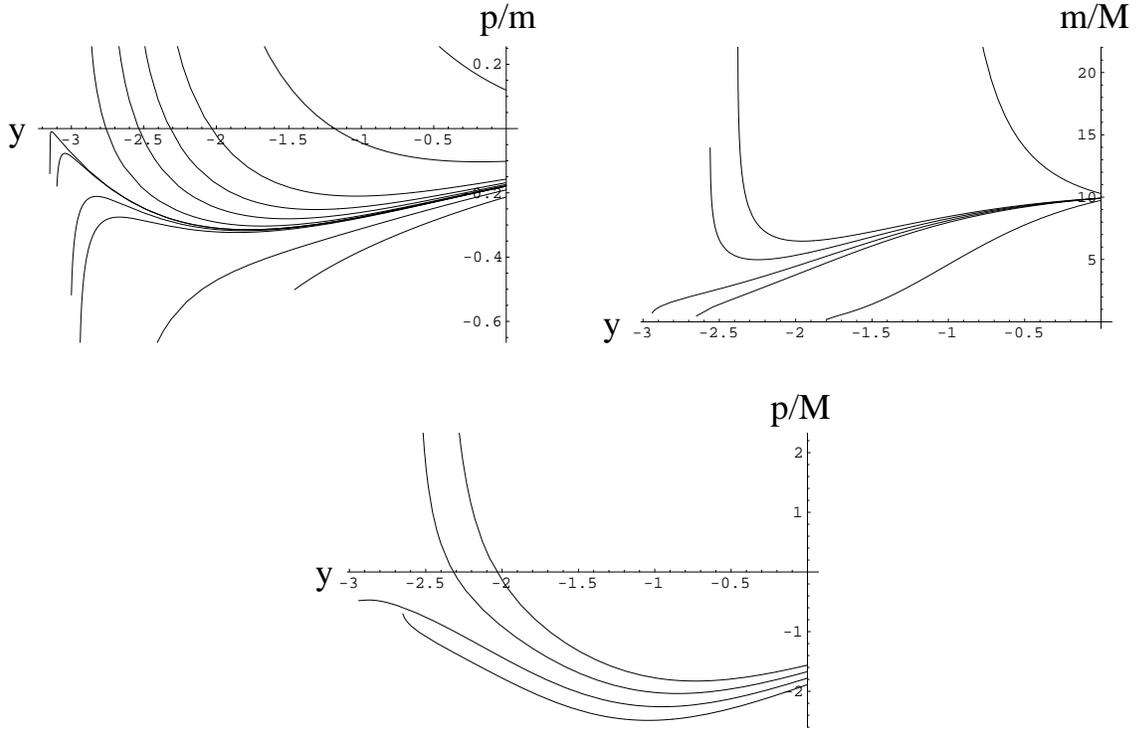}}
\caption{The evolution of the ratios $\rho/m, m/M$ and $\rho/M$ with $y$
as function of $\rho(y_{UV})$.}
\label{Fig5}
\end{figure}

From Fig. \ref{Fig4}, one can see that our solutions are all singular in
the IR.  
It is then  natural to ask whether they  correspond to
physical or unphysical flows.  
The expectation is that only the ridge
flow should be physical. Unfortunately we can explicitly check only the case 
$k<0$, where we can extract information from the IR
asymptotic behaviours
\bea
m & \sim & \mbox{const},\\
M & \sim& -\frac{3}{7} \log|y-y_0| + \mbox{const}, \\
  \rho &\sim & \frac{\sqrt{6}}{14} \log|y-y_0| +\mbox{const}, \\
\phi & \sim & \frac{1}{7} \log|y-y_0| + \mbox{const}. 
\eea{meep5}
Notice in particular the logarithmic divergence of the field 
$\phi$:  $\phi \sim A \log |y-y_0|$. Indeed for logarithmically
divergent  flows, the
criterion proposed in \cite{gub2} to select physically sensible solutions
seems to pick up flows with $A\ge 1/4$ \cite{pz}. This rules out the 
$k<0$ flows for which $A=1/7$.

For the ridge and the $k>0$ flows we can rely only on the 
numerical behaviour of
the potential and the superpotential, that, as observed above,
seem to allow only for one physical flow, the ridge.  
%
%\bea
%m &\sim& -\frac{3 \sqrt{2}}{4} \log|y-y_0| + const.,\\
%M &\sim&  -\frac{3}{8} \log|y-y_0| + const, \\
%\rho &\sim &  \frac{\sqrt{6}}{8} \log|y-y_0| + const,\\ 
%\phi &\sim&  \frac{1}{4} \log|y-y_0| + const.  
%\eea{meep7}
%  
%while for the $k<0$ flows we have for $y \rightarrow y_0$
%\beq
%\begin{array}{cccccccc}
% k < 0: &  m & \sim & const, \,\,\,\,\,\,\,\,\, & k > 0:  
%& m & \sim& -\frac{\sqrt{3}}{2} \log|y-y_0| + const,\\
%& M & \sim& -\frac{3\sqrt{2}}{14} \log|y-y_0| + const, & 
%& M &\sim&  const, \\
%
%
%&  \rho &\sim & \frac{\sqrt{6}}{14} \log|y-y_0| + const, & 
%&\rho &\sim& -\frac{\sqrt{6}}{7} \log|y-y_0| + const,\\ 
%&\phi & \sim & -\frac{1}{7} \log|y-y_0| + const, 
%& &\phi & \sim & -\frac{1}{7} \log|y-y_0| + const. \end{array}\\
%\eeq{meep5}
% 
%  

Finally let us consider the situation where the two masses $m$ and $M$
are of the same order. In field theory we expect to recover pure N=1 SYM. In
particular the vacuum state should evolve into the N=1 solution of
\cite{gppz3}. 
From the supergravity point of view the N=1 and the N=2 theories can be
distinguished by the scalars one has to turn on. As discussed in
\cite{gppz3}, in the N=1 solution the scalar $\rho$ is set to
zero. In this case, the supergravity mode corresponding to the mass
term for the scalar fields corresponds to the stringy mode only and it is therefore
not present in the 5$d$ supergravity Lagrangian.
Thus we expect that as we increase $M(y_{UV})$
the ridge line flow should smoothly move to the value $\rho = 0$.

\begin{figure}[ht]
\epsfxsize 15cm \centerline
{\epsffile{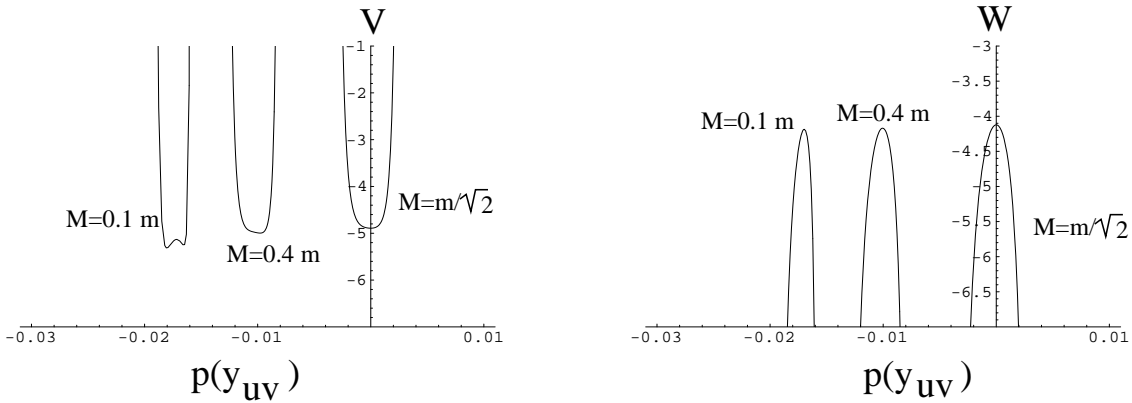}}
\caption{The evolution of the position of the 
critical curve with varying $M(y_{UV})$ from the N=2 to pure N=1 theory.
 }
\label{Fig7}
\end{figure}

This is indeed the case, as can be seen  in Fig. \ref{Fig7} where we have
plotted the 
potential and superpotential 
as a function of the UV values of $\rho$, in the deep IR
(small $y$), for 
three different values of the ratio $M(y_{UV})/m(y_{UV})$. 
$M(y_{UV})=m(y_{UV})/\sqrt{2}$ corresponds to pure
N=1 SYM and the
superpotential ridge has indeed moved to $\rho=0$.

\section{Discussion}

We have calculated the 5$d$ SUGRA scalar potential corresponding to fields
dual to scalar and fermion masses and a scalar VEV for the lightest
chiral multiplet in N=4 SYM softly broken in a cascade to N=2 and then
N=1 SYM. We have found numerical solutions of the resulting SUGRA
equations of motion and used the criteria that the potential must fall all along the flow, as suggested in \cite{gub2}, 
to distinguish physical flows. Interpreting these flows in terms of the
dual field theory  leads to a pleasing picture of the transition
between the N=2 flows of \cite{gub2, pw}  and the N=1 flows of
\cite{gppz3}. The N=2 theory has a moduli space and 
thus there is a class of flows in SUGRA which are physically
acceptable. The extremal curve has been identified with the singular
point on the moduli space. When a small mass is introduced 
to break the theory to N=1 SYM all these flows except the extremal flow
have their potential lifted in the IR which, according to the criteria above,
indicates they are unphysical. 
We thus see the unique vacuum of the N=1 theory emerging from the N=2
theory and, as expected from the field theory, the N=1 theory is pinned
at the singular point on the N=2 moduli space. As the two perturbing
masses (the one that sets the breaking N=4 to N=2 and the one that
further breaks to N=1) become  of the same order, leaving N=1 SYM in the IR,
the extremal flow is seen to smoothly 
move to the origin of the N=2 theory's moduli space, again in agreement
with field theory expectations. 

A number of issues remain for future investigation. 
The N=1 theory is
expected to have a gaugino condensate, represented in SUGRA by a
further scalar we have neglected in our discussion. Although the potential 
 is easy to compute, including this scalar would raise the dimension of
the parameter space 
of flows making   
analysis harder, so we prefer to leave it for future analysis. 
From the field theory we also 
expect to see monopole condensation and confinement
(in the field theory of the softly broken N=2 theory there are
$N^2-1$ different string tensions one might hope to see a SUGRA dual of
\cite{sw1}). 
The solutions we present
have an IR singularity so we could hope that the ridge flow 
solution exhibits confinement.
However 
we have no means to check the behaviour without explicit solutions. 
The usual methods one can apply in supergravity to check
confinement, such as the computation of a Wilson loop \cite{malda2}
 or the 
the evaluation of the spectrum of scalars \cite{w2,oz}, all require
at least the 
knowledge of the IR
asymptotic behaviour of the solution. 
In this respect it may be profitable to study the lift
of the 5$d$ SUGRA to the full 10$d$ theory where explicit solutions
may be more readily obtainable. The 10$d$ solutions would also
allow study of the running of the gauge coupling which would provide further
checks of our interpretation.

\vskip .2in
\noindent
{\bf Acknowledgements}\vskip .1in
\noindent
M.P. would like to thank A. Zaffaroni and G. Salam for very useful
discussions and comments.
M.P is partially supported by INFN and
MURST, and by the European Commission 
TMR program ERBFMRX-CT96-0045, 
wherein she is  associated to Imperial College, London, and by the PPARC
SPG grant 613. N.E is grateful for the support of a PPARC Advanced 
Fellowship.

\end{document}